\documentclass{emulateapj}
\usepackage{graphicx}

\shorttitle{Reconnection Outflow and CME Triggering in a Small Solar Eruption} 
\shortauthors{Reeves et al.} 

\begin{document}

\title{Direct Observations of Magnetic Reconnection Outflow and CME Triggering in a Small Erupting Solar Prominence}
\author{Katharine K. Reeves, Patrick I. McCauley and Hui Tian}
\affil{Harvard-Smithsonian Center for Astrophysics, 60 Garden St. MS 58, Cambridge, MA 02138}
\email{{\tt kreeves@cfa.harvard.edu}}

\begin{abstract}
We examine a small prominence eruption that occurred on 2014 May 1 at 01:35 UT and was observed by the {\it Interface Region Imaging Spectrometer} ( {\it IRIS}) and the Atmospheric Imaging Assembly (AIA) on the  {\it Solar Dynamics Observatory} ( {\it SDO}).  Pre- and post-eruption images were taken by the X-Ray Telescope (XRT) on  {\it Hinode}.  Pre-eruption, a dome-like structure exists above the prominence, as demarcated by coronal rain.  As the eruption progresses, we find evidence for reconnection between the prominence magnetic field and the overlying field.  Fast flows are seen in AIA and  {\it IRIS}, indicating reconnection outflows.  Plane-of-sky flows of ~300 km s$^{-1}$ are observed in the AIA 171 A channel along a potentially reconnected field line.     {\it IRIS} detects intermittent fast line-of-sight flows of ~200 km s$^{-1}$ coincident with the AIA flows.  Differential emission measure calculations show heating at the origin of the fast flows.  Post-eruption XRT images show hot loops probably due to reconfiguration of magnetic fields during the eruption and subsequent heating of plasma in these loops.   Although there is evidence for reconnection above the prominence during the eruption, high spatial resolution images from  {\it IRIS} reveal potential reconnection sites below the prominence.  A height-time analysis of the erupting prominence shows a slow initial rise with a velocity of ~0.4 km s$^{-1}$ followed by a rapid acceleration with a final velocity of ~250 km s$^{-1}$. Brightenings in  {\it IRIS} during the transition between these two phases indicate the eruption trigger for the fast part of the eruption is likely a tether-cutting mechanism rather than a break-out mechanism. \end{abstract}
\keywords{sun: flares, sun: coronal mass ejections, sun: activity}

\section{Introduction}
Reconnection is a fundamental process through which energy is released in solar eruptions.  Although the diffusion region where reconnection takes place is too small to be observed directly, the consequences of reconnection on the Sun occur on large scales that can be readily observed with current instrumentation.  The reconfiguration of magnetic fields due to reconnection is expected to cause Alfv{\'e}nic outflows from the reconnection region, sometimes resulting in shocks, and plasma heating is expected to occur due to these shocks or Ohmic dissipation of currents \citep[e.g.][]{Parker1957,Sweet1958,Petschek1964,PriestForbesBook}.    

Outflows due to reconnection during solar flares have been previously observed by solar imaging telescopes.  \citet{Savage2010} observed both sunward and anti-sunward flows eminating from the same location in a post-eruption plasma sheet using data from {\it Hinode}'s X-Ray Telescope \citep[XRT;][]{Golub2007}.  The speeds in this ``disconnection event'' were measured to be several hundred km s$^{-1}$.  Bi-directional outflows with speeds of several hundred km s$^{-1}$ have also been observed by the Atmospheric Imaging Assembly \citep[AIA;][]{Lemen2011} on the {\it Solar Dynamics Observatory} ({\it SDO}) during several more recent eruptions \citep{Savage2012b,Takasao2012,LiuW2013, Su2013}.  Supra-arcade downflows \citep[e.g.][]{McKenzie1999}, which are thought to be related to reconnection outflows, typically exhibit initial speeds of several hundreds of km s$^{-1}$ \citep{Sheeley2004,Savage2011,Warren2011,LiuW2013,LiuR2013}.

\begin{figure*}
\includegraphics[scale=0.62]{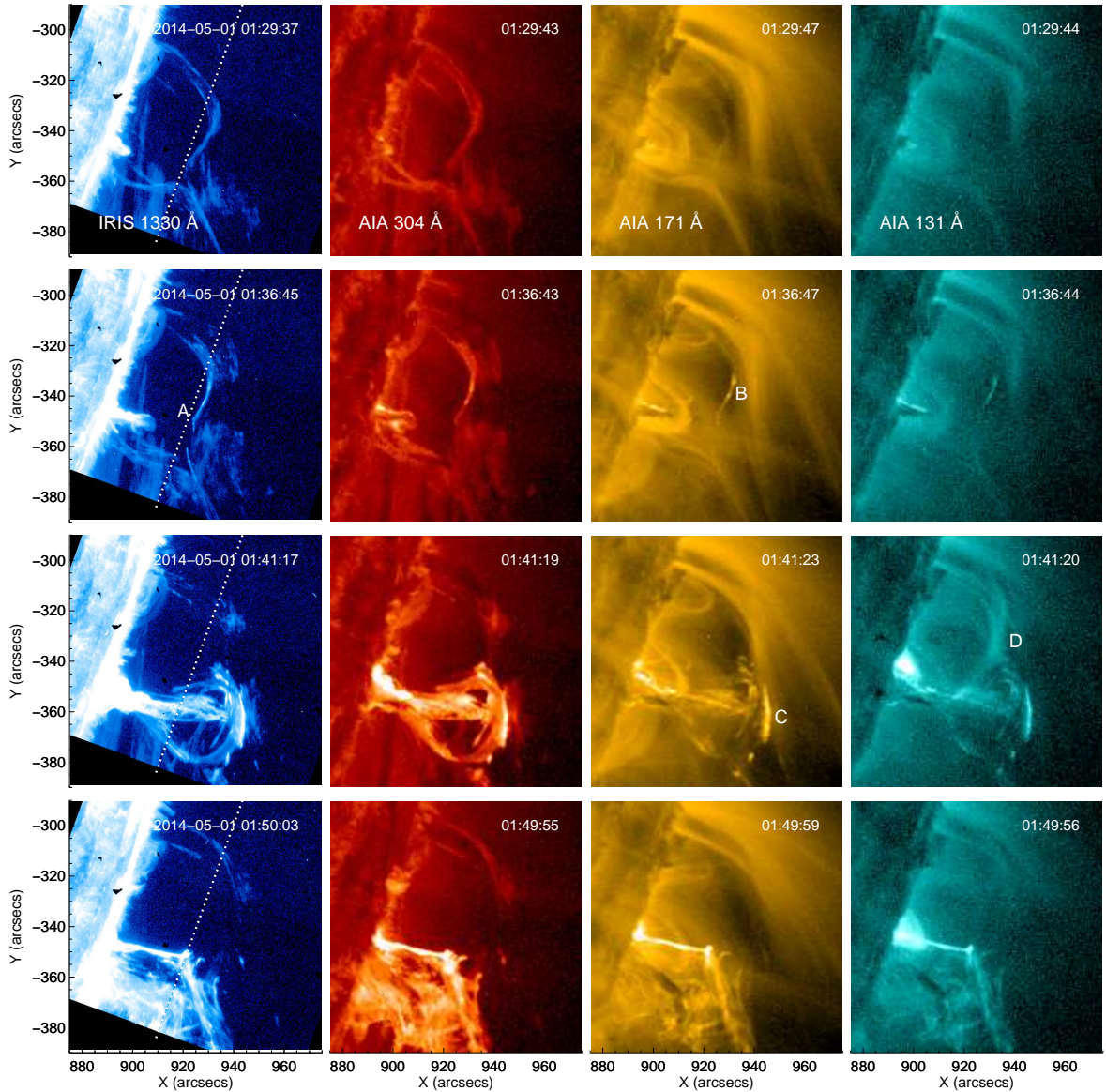}
\caption{\label{summary.fig}  {\it IRIS} 1330~\AA\ SJI images (left), AIA 304~\AA\ (second column), AIA 171~\AA\ (third column) and AIA 131~\AA\ (right) at several different times during the eruption.  The {\it IRIS} data has been rotated clockwise by 20 degrees. The white dotted line on the {\it IRIS} images shows the location of the slit.  A color version of this figure and an animation are included with the online material.}
\end{figure*}

Spectroscopic measurements of reconnection outflows have also been recorded.  \citet{Wang2007} measured enhanced wing emission indicative of fast flows with speeds of $\sim$900 -- 3500 km s$^{-1}$ in a flare observed with the Solar Ultraviolet Measurements of Emitted Radiation (SUMER) spectrometer on the {\it Solar and Heliospheric Observatory} ({\it SoHO}) mission.   SUMER has also recorded strong blue-wing enhancements of $\sim1000$  km s$^{-1}$ associated with supra-arcade downflows \citep{Innes_b2003,Innes_a2003}. \citet{Hara2011} measured a Doppler shift corresponding to 200 -- 400 km s$^{-1}$ in the Fe XXIV and Ca XVII lines during a flare measured by the Extreme ultraviolet Imaging Spectrometer (EIS) on the {\it Hinode} mission, which they interpret as an anti-sunward reconnection outflow.    Recently, \citet{Tian2014} observed red shifts corresponding to reconnection outflows in the Fe XXI line during a flare observed by the {\it Interface Region Imaging Spectrometer} ({\it IRIS}).

In most of the eruptions described above, the observed reconnection outflow occurs in a current sheet-like structure above a set of cuspy flare loops, as in the standard CSHKP \citep{Carmichael1964,Sturrock1968,Hirayama1974,Kopp1976} model of solar flares.  In this paper, we examine the evidence for reconnection in an altogether different location.  We find evidence for reconnection between an erupting prominence and the pre-existing overlying magnetic field in an event that occurred on 2014 May 1 at about 01:35 UT.  We also examine the possible triggers of the eruption.  The outline of the paper is as follows: the observations are presented in \S\ 2, our results are presented in \S\ 3, and discussion and conclusions are presented in \S\ 4.

\section{Observations \label{observations.sec}}

A small eruption occurred at the limb of the Sun on 1 May 2014, starting at 1:35 UT.  This eruption was observed by   {\it IRIS} \citep{DePontieu2014} with the 1330 \AA\ and 2796 \AA\ slit jaw imagers (SJIs) using a spatial pixel size of $\sim$0.166\arcsec\ per pixel, a field of view of 120\arcsec$\times$120\arcsec\ and a cadence of about 20 seconds for each passband.   The spectral resolution of {\it IRIS} in this observation is ~1.3 m\AA\ per pixel in the far ultraviolet (FUV) bandpass.  We use level 2 data, which has been dark corrected, flat fielded, and geometrically corrected.  Spectra are observed along a slit centered at X=931.05\arcsec, Y=-321.26\arcsec\ throughout the event.   The entire slit is located off-limb in this observation, meaning that the cool chromospheric lines usually available for calibration are not present in the spectra.  Thus we set the rest wavelength in the {\it IRIS} spectra by averaging the spectra over space and time, assuming that the centroid of the average profile (fit by a Gaussian function) is located at zero velocity.

Full sun images from AIA \citep{Lemen2011} are available with a resolution of $\sim$0.6\arcsec\ per pixel and a cadence of 12 seconds for the EUV passbands.  The AIA data are deconvolved with the point spread function in each filter using the routine {\tt aia\_deconvolve\_richardsonlucy} and then processed using the {\tt aia\_prep} routine, which de-rotates the images from the different AIA telescopes, aligns them, and gives them all the same plate scale. These AIA data processing routines are available in the SolarSoft Ware  \citep[SSW;][]{Freeland1998} package, available as an extension to IDL.

Summary images of the eruption observed by the {\it IRIS} 1330 \AA\ slit jaw imager (SJI) and the AIA 304 \AA, 171 \AA, and 131 \AA\ bandpasses are shown in Figure \ref{summary.fig}. The {\it IRIS} spacecraft was rolled by 20 degrees during the eruption, and the SJI images have been rotated to match the AIA field of view.  After rotating the {\it IRIS} images, they were aligned by eye with the AIA 304 \AA\ images by shifting the {\it IRIS} images.

As shown in the top row of Figure \ref{summary.fig}, coronal rain is observed in the {\it IRIS} 1330 \AA\ SJI image and the AIA 304 \AA\ image prior to the eruption, forming a dome shape over the eruption site.  The dome-shaped structure is reminiscent of a fan and spine type null point magnetic field configuration \citep[e.g.][]{Pontin2013, Sun2013, Masson2014}.   The spine part of the structure is also clearly visible in the AIA 171 \AA\ image and faintly visible in the AIA 131 \AA\ image.

The second row of Figure \ref{summary.fig} shows images around 01:36:45 UT, a time early in the evolution of the eruption.  As the small filament below the dome loses stability and starts to erupt, the coronal rain material in the overlying structure in the {\it IRIS} 1330 \AA\ image brightens and forms a distinct kink shape, marked with the letter `A' in Figure \ref{summary.fig}.  A brightening is also seen in the overlying structure in the AIA 304 \AA\ image, as well as a suggestion of the same kink shape seen in the {\it IRIS} image, but the coarser spatial resolution of the AIA image makes this shape less obvious.  There is a small linear structure that becomes bright in the AIA 171 \AA\ image that was not present before, marked with the letter  `B' in Figure \ref{summary.fig}.  This structure is also faintly visible in the AIA 131 \AA\ image taken a few seconds earlier.   In the movie supplied with the online material, it can be seen that this brightening travels along the dome structure in the few minutes after these images.

The third row of Figure \ref{summary.fig} shows the continuation of the event at around 01:41:17 UT as the erupting filament visibly pushes against the overlying dome structure.   All four images in this row show a brightening of the plasma where the material from the erupting filament merges with the plasma from the overlying dome structure, marked with the letter `C' in the AIA 171 \AA\ image.  The AIA 131 \AA\ image at this time shows a diffuse brightening at the same location where the small bright linear structure appeared at the previous time, labeled with the letter `D' in Figure \ref{summary.fig}.   There is no corresponding emission in the AIA 171 \AA\ image at the same location.

The final row of Figure \ref{summary.fig} shows the dissipation of the erupting filament at about 01:50:03 UT.  The remnants of the erupting filament can be seen in all four bandpasses, as well as a bright linear structure.  No coronal mass ejection (CME) was recorded for this time and location in the CACTus \citep[Computer Aided CME Tracking;][]{Robbrecht2009} CME catalog and visual inspection of images from the Large Angle and Spectrometric Coronagraph Experiment (LASCO) on {\it SoHO} do not show any evidence for a CME, so it seems that this eruption was ultimately a failed CME.

\begin{figure}
\includegraphics[scale=0.68]{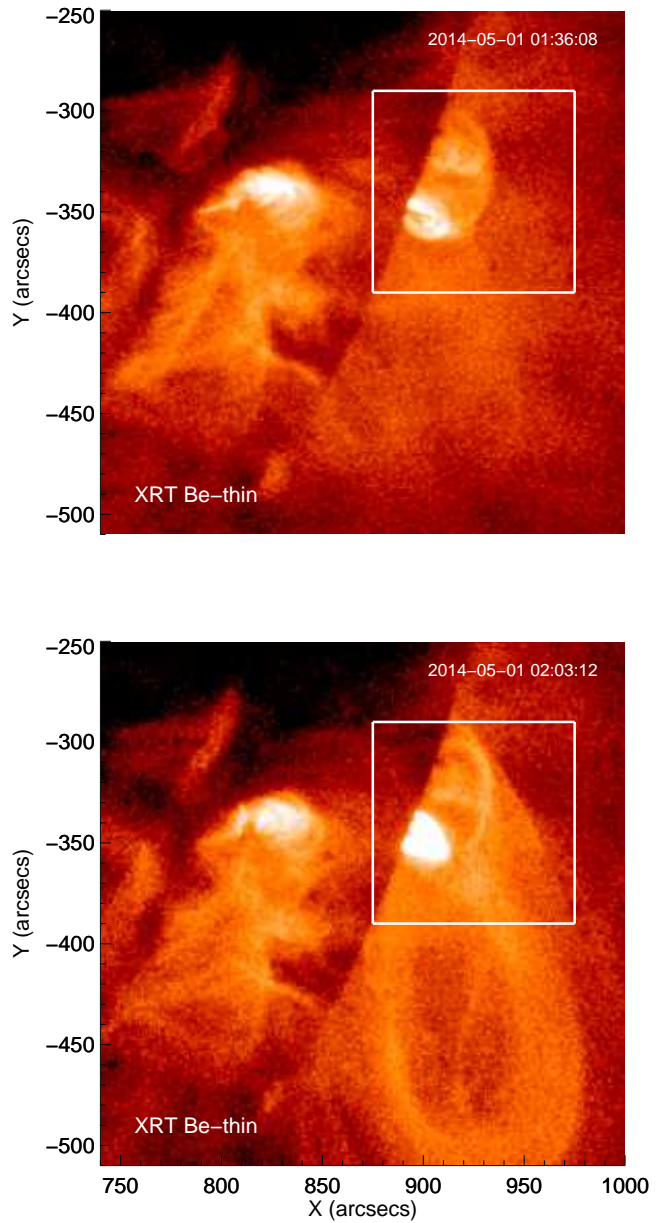}
\caption{\label{xrt_summary.fig}  XRT images before and after the eruption.  The box shows the field of view of the images in Figure \ref{summary.fig}.  (A color version of this figure is available in the online journal.)}
\end{figure}

The {\it Hinode} satellite \citep{Kosugi2007} was in eclipse during the eruption itself, but there are images from XRT \citep{Golub2007,Kano2008} in the Be-thin filter at a spatial resolution of 1.0286\arcsec\ per pixel just before the eruption at 01:36:08 UT and just after the eruption at 02:03:12 UT.   Additionally, there are sets of multi-filter XRT data consisting of the Al-poly, C-poly, Be-thin, Be-med, Al-med and Al-thick filters before and after the eruption at about 00:58 UT and 02:07 UT, respectively.    We use these filters \citep[except for the Al-poly and C-poly filters, which suffer from contamination issues, see][]{Narukage2011} to investigate the thermal properties of the plasma before and after the eruption.   XRT images are processed using the SSW routine {\tt xrt\_prep}, which subtracts the dark current and removes the CCD bias and telescope vignetting \citep[see][for details]{Kobelski2014}.  The XRT data is also deconvolved with   
an appropriate point spread function. 

 XRT Be-thin images just before and just after the eruption are shown in Figure \ref{xrt_summary.fig}.  In the image taken before the eruption at 01:36:08 UT,  there is a bright feature in the southern part of the active region that corresponds to the erupting filament seen a few seconds later in the {\it IRIS} and AIA images.  In the image taken after the eruption at 02:03:12 UT, there is a large loop that extends to the south of the active region that was not present in the XRT image taken just before the eruption.  In this image, a bright set of flare loops can be easily seen in the location where the filament was prior to the eruption.

\begin{figure*}
\includegraphics[scale=0.55]{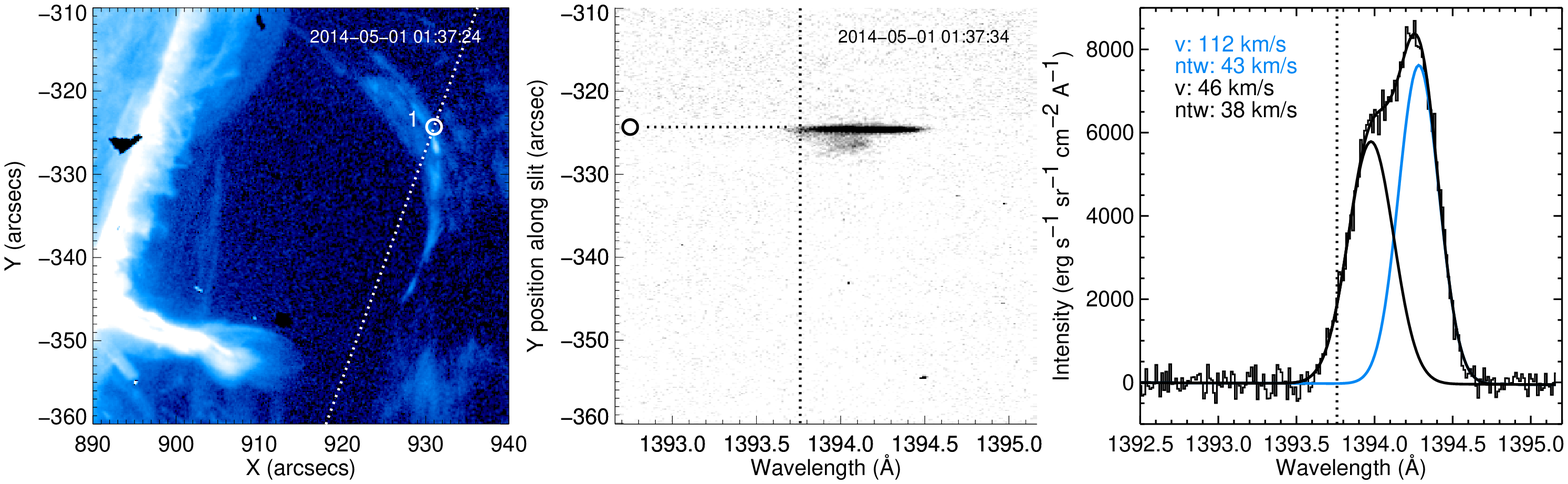}
\includegraphics[scale=0.55]{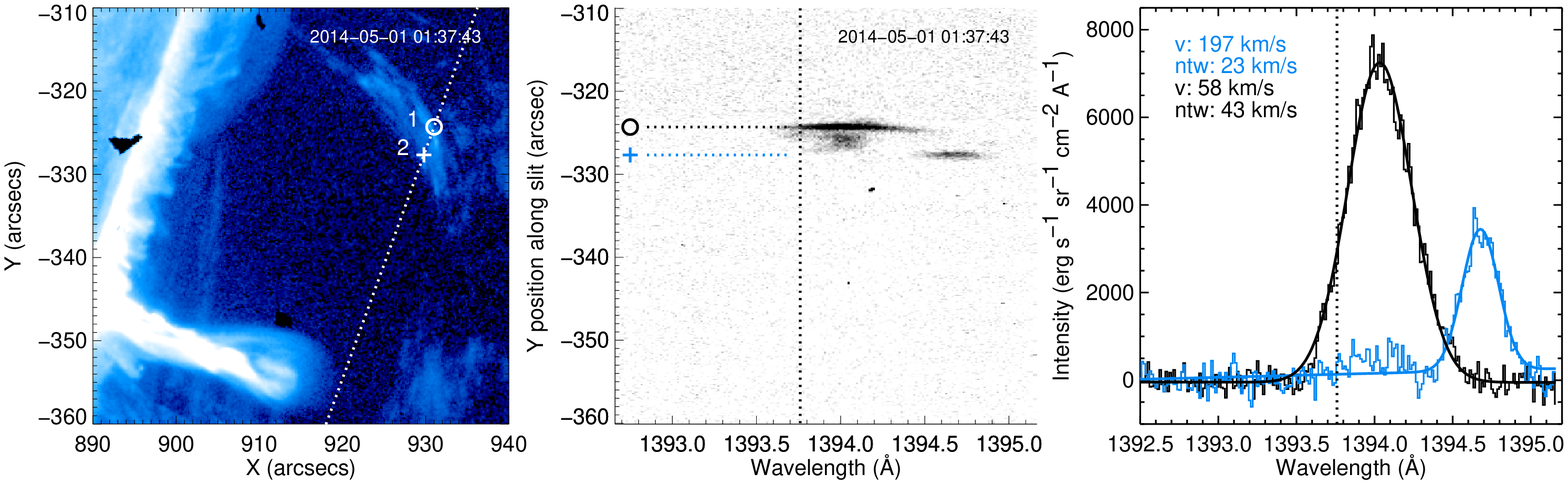}
\includegraphics[scale=0.55]{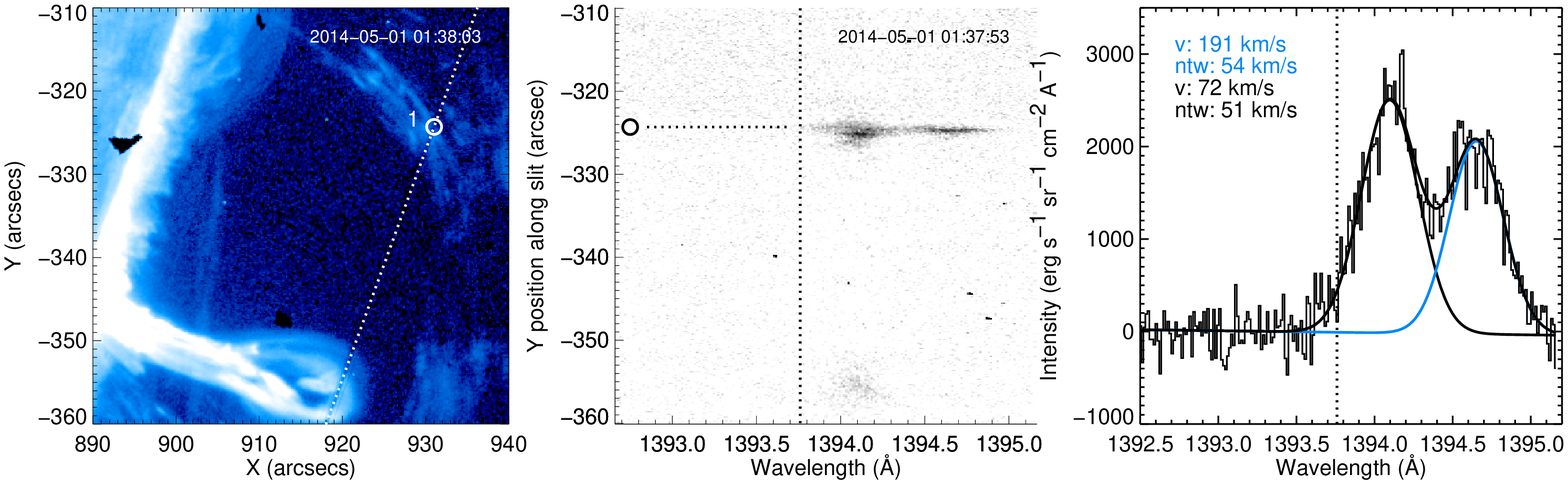}
\caption{\label{iris_redshift.fig}  {\it IRIS} 1330 \AA\ slit jaw images (left column), spatial images of the portion of the {\it IRIS} FUV spectra encompassing the Si IV 1393.76 line (middle column), and plots of the Si IV line profiles (right column) for several different times.  The Locations 1 and 2 are marked by a circle and a cross, respectively, on the images in the first two columns.  The location of the {\it IRIS} slit is marked with a dotted line in the images in the left column, and the Si IV rest wavelength is marked with a dotted line in the middle and right columns.  In the top and bottom rows, the spectra at Location 1 are fit with two Gaussians.  In the middle row, the spectra at both Locations 1 and 2 are fit with single Gaussians.  The faster components of the IRIS line profiles are colored grey (blue in the online figure).   Listed on the plots of the spectra are the calculated velocity (v) and the non-thermal width (ntw) of the lines.
(A color version of this figure is available in the online journal.)}
\end{figure*}

\section{Results \label{results.sec}}

\subsection{Flows and Non-thermal Broadening from {\it IRIS}}

The brightening in the coronal loop at the location marked by an `A' in the {\it IRIS} 1330 \AA\ SJI image in Figure \ref{summary.fig} travels along the loop defined by the coronal rain until it crosses the {\it IRIS} slit starting at 01:37:34 UT, when strong redshifts ($>$ 100 km s$^{-1}$) begin to be seen in the {\it IRIS} Si IV line at 1393.73 \AA, as shown in the top row of Figure \ref{iris_redshift.fig}.  Prior to this time, in the region where the coronal rain crosses the {\it IRIS} slit, red shifts of about  65 km s$^{-1}$ and non-thermal widths of about 36 km s$^{-1}$ are observed.  The strong red shifts are seen at two different locations along the {\it IRIS} slit, centered at approximately (x,y) = (931.01\arcsec,-324.31\arcsec), which we will refer to as Location 1,  and (x,y) = (929.79\arcsec,-327.67\arcsec), which we will refer to as Location 2.  Spectra from Location 1 and Location 2 are averaged spatially over six and five {\it IRIS} pixels, respectively.  Location 1 is marked with a circle and Location 2 is marked with a cross on the {\it IRIS} SJI images in Figure \ref{iris_redshift.fig}.

The top row of Figure \ref{iris_redshift.fig} shows the spectrum of the Si IV line at 1393.76 \AA\ at 01:37:34 UT, when the first high velocity red shifts become visible at Location 1.   Figure \ref{iris_redshift.fig} also shows an image from the {\it IRIS} 1330 \AA\ SJI as well as the image of the spectrum on the CCD. The spectrum from this time is fit with a double Gaussian function, indicating two distinct velocity components along the line of sight.  The highly red-shifted component has a speed of 112 km s$^{-1}$. The other component is red shifted with a speed of ~58 km s$^{-1}$.  

We also measure the line widths of the two velocity components in order to determine their non-thermal widths, which are an indication of  unresolved motions along the line of sight, often due to turbulence or wave motions. The observed line width is given by:
\begin{equation}
\Delta\lambda = \frac{\lambda_{0}}{c}\sqrt{\frac{2kT}{m} + \xi^2 + \sigma_I^2},
\end{equation}
where $\lambda_0$ is the rest wavelength, $c$ is the speed of light, $k$ is Boltzmann's constant (1.38$\times10^{-23}$m$^2$ kg s$^{-2}$ K$^{-1}$), $m$ is the mass and $T$ is the temperature of the emitting ion, $\xi$ is the non-thermal width, and $\sigma_I$ is the instrumental width.  We express the components of the line width as 1/e widths.  We assume an instrumental broadening of 4.1 km s$^{-1}$, as in \citet{TianScience2014}, and a $T$ equal to the formation temperature of Si IV ($10^{4.9}$ K).  Using these assumptions, we find that for the high-speed component in the top row of Figure \ref{iris_redshift.fig}, the non-thermal width is 43 km s$^{-1}$, while for the less red-shifted component, the non-thermal width is 38 km s$^{-1}$, which are larger than the typical non-thermal widths of Si IV observed in the quiet Sun \citep[$\sim$ 20 km s$^{-1}$;][]{TianScience2014}.

The fastest red-shifted component comes 10 s later at 01:37:43, and it is centered at Location 2, which is closer to the surface of the Sun by about 3.3\arcsec\ than Location 1.  The spectrum of the Si IV 1393.73 \AA\ line at both Location 1 and Location 2 at this time are shown in the middle row of Figure \ref{iris_redshift.fig}. The highly red-shifted component at Location 2 has a speed of 197 km s$^{-1}$. At Location 1,  the same location where the highly redshifted component was 10 seconds previously, the Si IV consists of a single component Gaussian with a red shift of about 58 km s$^{-1}$, similar to velocities seen at that location prior to the eruption.  The high-speed component at Location 2 has a non-thermal width of 23 km s$^{-1}$, while the slower component at Location 1 has a non-thermal width of 43 km s$^{-1}$.

The bottom row of Figure \ref{iris_redshift.fig} shows the Si IV spectrum from Location 1 at  01:37:53 UT.  At this time there are clearly again two components to the profile, and we use a double Gaussian function to fit the spectral profile.  The two components consist of a fast one red-shifted to about 191 km s$^{-1}$ and a slower one red-shifted to about 72 km s$^{-1}$.   Both of these components have similar non-thermal widths -- the faster component has a non-thermal width of 54 km s$^{-1}$, while the slower component is 51 km s$^{-1}$.

\subsection{Flows from AIA \label{aia_flows}}

In the AIA 171 \AA\ channel, a steady bright flow is seen (marked with a `B' in Figure \ref{summary.fig}) at the same location as the brightenings and fast flows that we see in the {\it IRIS} SJI images.  Since the coronal rain is not visible in the 171 \AA\ channel like it is in the {\it IRIS} SJI images and the AIA 304 \AA\ images,  we use this channel to determine the plane-of-sky velocity of the flows associated with the eruption.

\begin{figure*}
\includegraphics[scale=0.4]{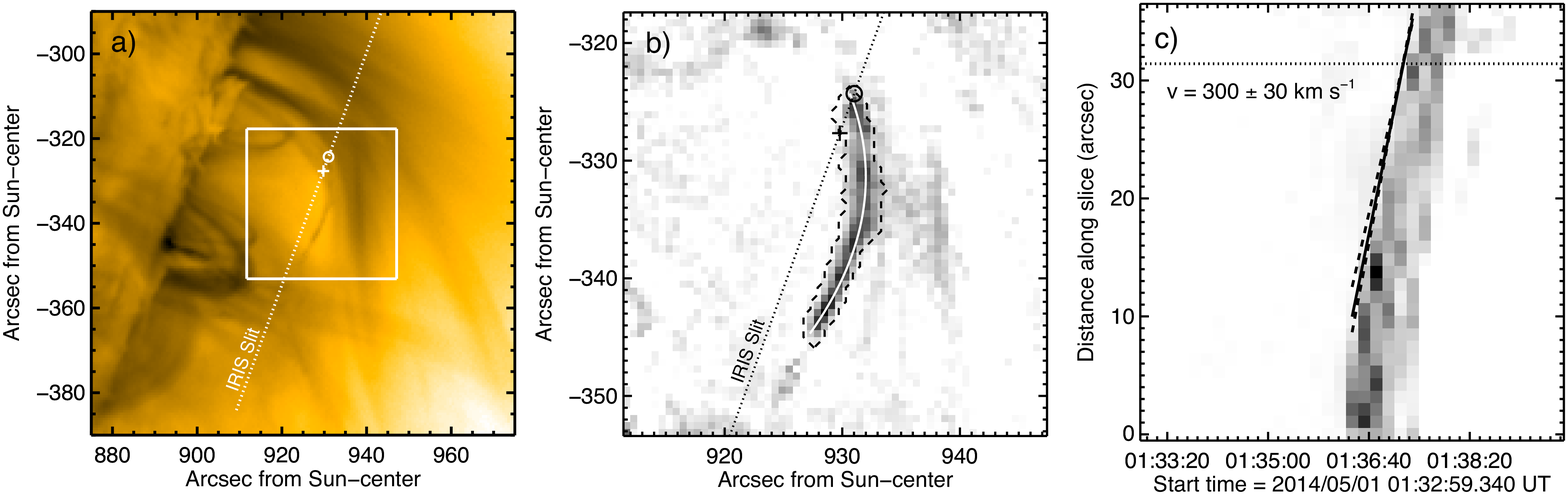}
\caption{\label{AIA-velocity.fig} Panel a): A sum of 12 AIA 171 \AA\ images starting at 01:36 UT. Panel b): Background subtracted sum, with a field of view as indicated by the white box in Panel a.  Panel c): Height-time plot of the motion along the slice defined by the white arc in Panel b.  In all the images, the location of the {\it IRIS} slit is given by a dashed line, and Locations 1 and 2 are marked by a circle and a cross, respectively, as in Figure \ref{iris_redshift.fig}.  A color version of this figure and an animation are included with the online material. }
\end{figure*}

To measure the plane-of-sky velocity from AIA 171 \AA\ images, we first define the spatial boundaries of the plasma flow. Figure \ref{AIA-velocity.fig}a shows the sum of twelve 171 \AA\ images taken over 2.5 minutes after the flow appears in this channel at 01:36 UT. The white box denotes the field-of-view of Figure \ref{AIA-velocity.fig}b, which shows the same sum subtracted by an earlier background. This image is thresholded to produce the dashed contour, and a parabolic arc is fit to the pixels within the contour to produce the white slice. 

Emission along the slice as it evolves in time, taken from row averages within the dashed contour, is shown in Figure \ref{AIA-velocity.fig}c. Each column represents an individual base-difference (background-subtracted) image taken after 01:33 UT. See the online material for a corresponding movie. The leading edge of this space-time plot is extracted using the Canny edge detection algorithm \citep{Canny1986} and fit with a line, the slope of which gives the velocity along the arc. The parameters of the edge detection are varied to obtain the three linear fits shown in Figure \ref{AIA-velocity.fig}c, which indicate a plane-of-sky velocity of 300 $\pm{30}$ km s$^{-1}$. 

The location of the {\it IRIS} slit is indicated in Figure \ref{AIA-velocity.fig} by a dashed line, and the locations of the fast {\it IRIS} flows are marked with a cross and a circle, as in Figure \ref{iris_redshift.fig}.  Both of the fast {\it IRIS} flows occur on the edge of the bright material that constitutes the flow in the AIA 171 \AA\ channel, with the more non-thermally broadened flow (circle) on the edge towards the external field and the less non-thermally broadened flow (cross) on the edge towards the erupting filament.

\subsection{Thermal Analysis}

 Reconnection of magnetic fields often includes heating of the plasma as part of the energy release.  In order to determine how the thermal energy in the plasma changes as a result of the erupting prominence, we calculate differential emission measures (DEMs) at a few key locations during the eruption. 

To calculate the DEMs, we use the algorithm  {\tt xrt\_dem\_iterative2}, written by Mark Weber \citep{Weber2004, Golub2004, Cheng2012}, which is available in the SSW database.  This algorithm works by adjusting a series of spline knots in the DEM solution until the $\chi^2$ between the measured and calculated intensities is minimized.  The algorithm has been shown to do a good job at reproducing model DEMs \citep[see the appendix of][]{Cheng2012}.  \citet{SchmelzKashyap2009, SchmelzSaar2009} have shown that {\tt xrt\_dem\_iterative2} provides solutions that are in good agreement with the Markov Chain Monte Carlo (MCMC) method developed by \citet{KashyapDrake1998}, and \citet{HannahKontar2012} have similarly found good agreement with their regularized inversion DEM method.  Errors on the DEMs are determined by performing 100 Monte Carlo simulations, where each Monte Carlo solution is a DEM calculated using the measured intensities varied by a normally distributed random error.  DEMs are calculated using all six Fe-dominated AIA EUV channels and, when available, the XRT Be-thin, Be-med, Al-med and Al-thick filters.  We do not use the XRT Al-poly and C-poly filters in the DEMs because they are more susceptible to contamination issues \citep{Narukage2011,Narukage2014}.

\begin{figure*}
\includegraphics[scale=0.6]{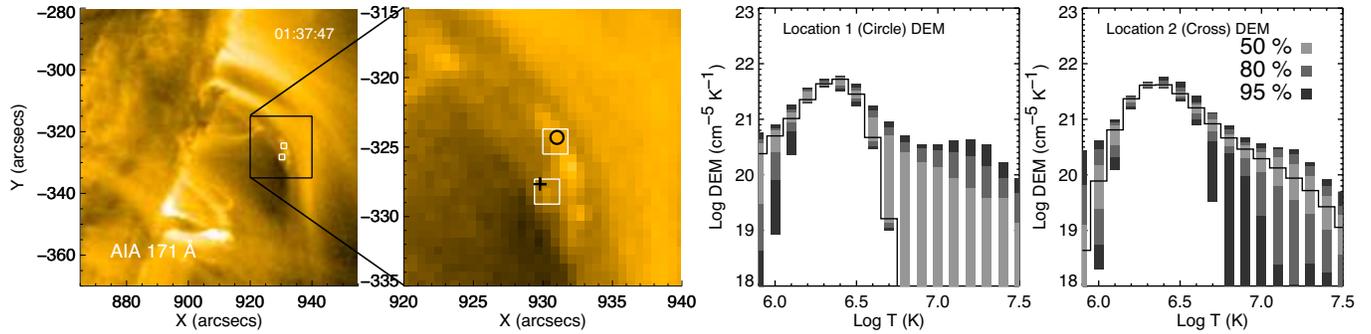}
\caption{\label{dem_outflows.fig} Images and DEMs at about 01:37:47 UT, the time at which the fastest flows were observed in IRIS.  The left two panels show an AIA 171 \AA\ image, with an overview on the far left and a smaller field of view in the center left panel.  The locations of the IRIS flows are marked with a cross and a circle on the center left image. White boxes indicate the region of interest where intensities were used to calculate DEMs.  The center right panel shows the DEM of the area around Location 1, marked with a circle, and the rightmost panel shows the DEM of the area around Location 2, marked with a cross. Both DEM plots include the 100 Monte Carlo solutions, where dark grey, grey and light grey boxes encompass 95\%, 80\% and 50\% of the Monte Carlo solutions, respectively. (A color version of this figure is available in the online journal.)}
\end{figure*}

\begin{figure*}
\includegraphics[scale=0.6]{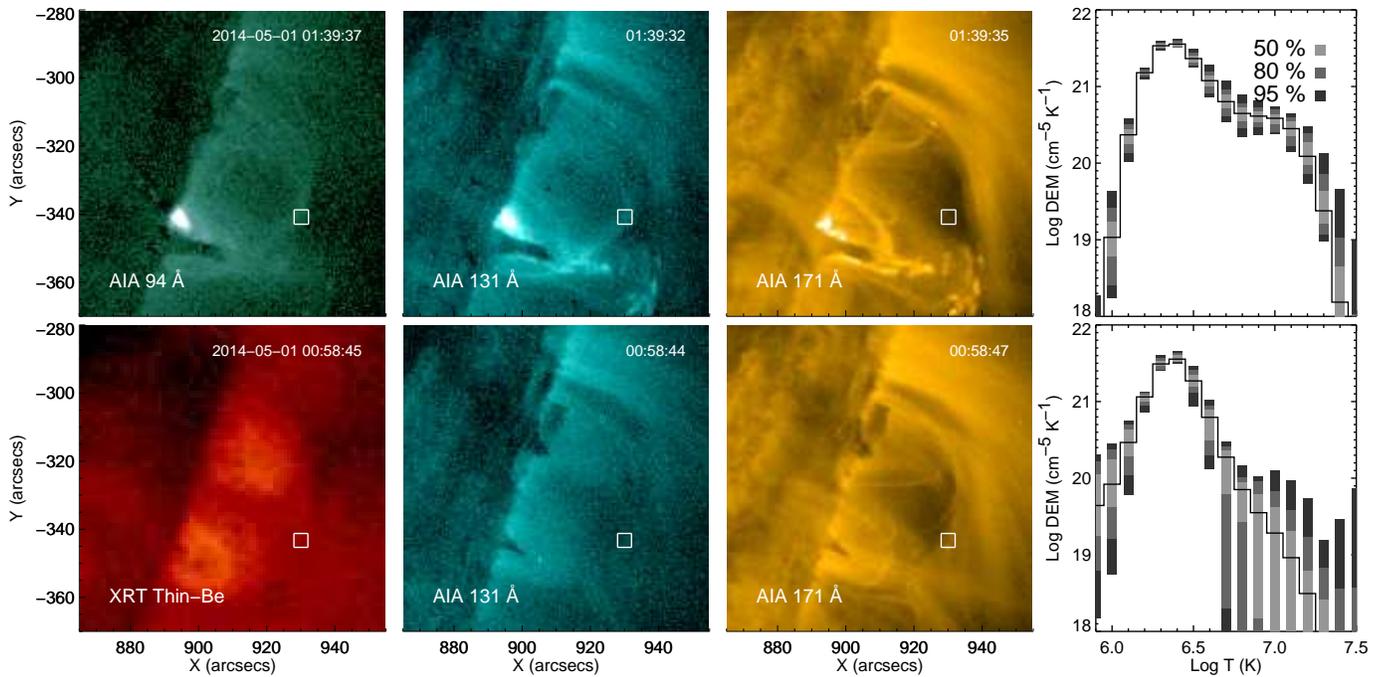}
\caption{\label{dem_small_loop.fig} Top row: Images and DEM for 01:39 UT, during the eruption.   Bottom Row:  Images and DEM for 00:58 UT, before the eruption. Images are scaled the same on the top and the bottom rows.  The DEMs are calculated by averaging the pixels in the white boxes marked on the images.  The solid lines in the DEM plots show the DEM calculated from the averaged intensities.  Dark grey, grey and light grey boxes on the DEM plot encompass 95\%, 80\% and 50\% of the Monte Carlo solutions, respectively. (A color version of this figure is available in the online journal.)}
\end{figure*}

The temperature response functions for the two telescopes used in the DEM calculations are from the latest versions currently available.  For AIA, the temperature responses used are Version 6 from the {\tt aia\_get\_reponse} program, using the ``chiantifix,'' ``evenorm'' and ``noblend'' keywords. The ``chiantifix'' keyword applies an empirical correction to the AIA 94 \AA\ response function to account for missing lines in the CHIANTI spectral package, and the ``evenorm'' keyword normalizes the AIA temperature responses to the output of the EUV Variability Experiment (EVE) on {\it SDO} \citep{Boerner2014}.  The ``noblend'' keyword removes the blending of the 131 \AA\ and the 335 \AA\ channels in the response functions, which has been previously found to provide better fits to DEM data \citep{McCauley2013,Hanneman2014}.  For XRT, the temperature responses are calculated with the calibrations outlined in \citet{Narukage2014} and using the same solar spectrum as is used for the AIA response functions.  For both telescopes, we use the date-dependent versions of the temperature response functions, which account for any loss in sensitivity of the instrumentation as a function of time.

 We first calculate DEMs at locations where IRIS observes fast line-of-sight flows at 01:37:43 UT (see Figure \ref{iris_redshift.fig}).  These DEMs are shown in Figure \ref{dem_outflows.fig}, and they are calculated by averaging the pixels in the white boxes.  The positions of the IRIS fast flows are demarkated by a cross and a circle, as in previous figures.  The DEMs for both locations have a peak at about  log T = 6.4 (2.5 MK).  The DEM for Location 2 has a long tail at higher temperatures, while the DEM for Location 1 falls off rapidly.  The Monte Carlo simulations show that for both of the DEMs, the peak is fairly certain, but there is a wide range of possible values surrounding the high-temperatures, making it difficult to say with certainty if there is a high-temperature component to the plasma in either case.  Unfortunately there are no XRT data at this time, which would help to constrain the higher temperatures.

A few minutes later, at 01:39:35 UT, there is a brightening observed in the AIA 131 \AA\ intensity at the location where the flows in AIA originate (marked with the letter `D,'  in the third row of Figure \ref{summary.fig}), without a corresponding brightening in the 171 \AA\ channel.   Because of the hot component in the 131 \AA\ response function \citep[see][for thorough discussions of the AIA temperature response functions]{Boerner2012, Boerner2014}, this difference is often an indication that hot ($\sim$10 MK) plasma is present.    

Figure \ref{dem_small_loop.fig} shows AIA 94, 131 and 171 \AA\ images and a DEM at around 01:39 UT, when the aforementioned intensity enhancement is just starting to appear in the 94 and 131 \AA\ channels (see the on-line movie accompanying Figure \ref{summary.fig}).  The DEM is calculated using intensities averaged over the white box plotted on the images.   There are no XRT data for this time, so only the six Fe-dominated AIA EUV filters are used in calculating the DEMs.   Figure \ref{dem_small_loop.fig} also shows the DEM for a corresponding box at 00:58 UT, prior to the eruption.  We choose this time because there is a full set of XRT filters available, which are used in addition to the AIA filters to calculate the DEM.  The DEM at 01:39 UT has two components, a low-temperature component that peaks at about log T = 6.3 (2 MK) and a high-temperature component that peaks at around log T = 7.0 (10 MK).   The DEM calculated for the time before the eruption does not show a similar high temperature component within the margin of error provided by the Monte Carlo runs.

\begin{figure*}
\includegraphics[scale=0.6]{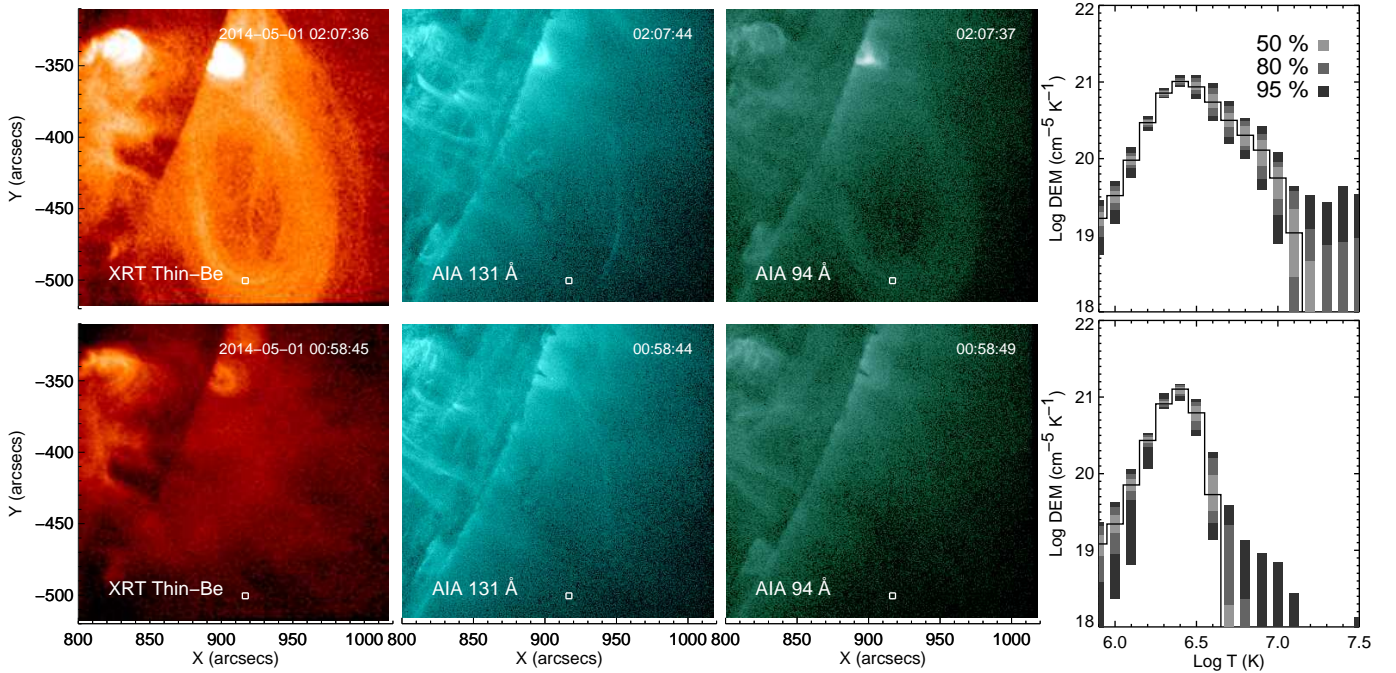}
\caption{\label{dem_big_loop.fig} Top row: Images and DEM for 02:07 UT, during the eruption.  Bottom Row:  Images and DEM for 00:58 UT, before the eruption.  Images are scaled the same on the top and the bottom rows.  The DEMs are calculated by averaging the pixels in the white boxes marked on the images.  The solid lines in the DEM plots show the DEM calculated from the averaged intensities.  Dark grey, grey and light grey boxes on the DEM plot encompass 95\%, 80\% and 50\% of the Monte Carlo solutions, respectively. (A color version of this figure is available in the online journal.)}
\end{figure*}

After the eruption, at 02:03 UT, there is a large loop seen in XRT images that does not exist before the eruption, as shown in Figure \ref{xrt_summary.fig},  indicating that some heating has taken place on this loop as well.  Figure \ref{dem_big_loop.fig} shows XRT/Be-thin, AIA/94, and AIA/131 \AA\ images and a DEM for 02:07 UT, a time after the eruption when a full set of XRT images is available.  During this time, the large loop that extends southward of the erupting loop that is shown in Figure \ref{xrt_summary.fig} is still clearly visible in the XRT images.  The only AIA filter where the loop is visible is the 94 \AA\ channel.  It is not visible in either the 335 \AA\ channel (not shown) or the 131 \AA\ channel.  

 As in Figure \ref{dem_small_loop.fig}, the DEM is calculated using averaged intensities in the white box.  We show a DEM from near the loop top, but DEMs from other locations are similar.  The DEM from 02:07 shows a dominant component peaked at log T = 6.4 (2.5 MK), but it also has a broad wing at higher temperatures.  This wing is not present in the pre-eruption DEM calculated at the same location, also shown in Figure \ref{dem_big_loop.fig}.  If we fit a double-Gaussian function to the DEM, we find that the wing is caused by a weak component peaked at about log T = 6.6 (4 MK).   This secondary component provides an explanation for why the big loop is weakly visible in the 94 \AA\ channel, but not the 335 \AA\ or the 131 \AA\ channels.  The peak temperature of log T = 6.6 is in a region of the 131 \AA\ response function where there is very little sensitivity.  The 335 \AA\ response function is sensitive to plasma at log T = 6.6 but is much more sensitive to plasma at the temperature of the dominant component of the DEM, log T = 6.4, so the background corona is likely making it difficult to see the big loop in this filter.  The 94 \AA\ temperature response reaches a minimum right around log T = 6.4, so the background corona does not contribute very much at all to the emission in this filter, thus it is able to weakly image the hotter component.  The images in Figure \ref{dem_big_loop.fig} also show the value of the XRT instrument in identifying hot plasma, since the big loop is clearly identified in the XRT image but only very faintly seen in one of the AIA filters.

\subsection{Eruption Triggering}
 
\begin{figure*}
\includegraphics[scale=0.65]{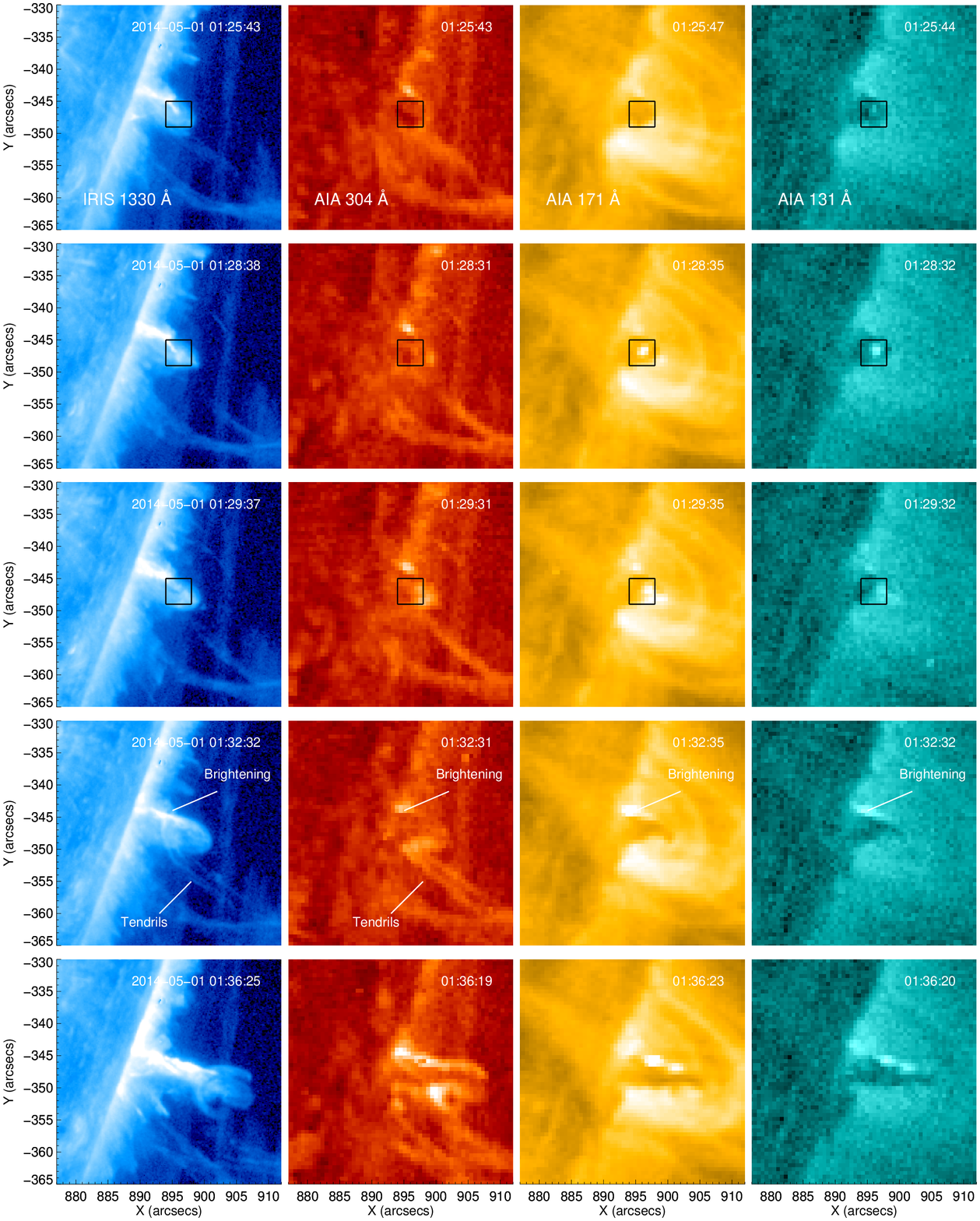}
\caption{\label{triggering.fig}  Close up images from the {\it IRIS} 1330~\AA\ SJI image (left), AIA 304~\AA\ (second column), AIA 171~\AA\ (third column) and AIA 131~\AA\ (right) showing the initiation of the eruption.  The {\it IRIS} data has been rotated clockwise by 20 degrees. A color version of this figure and an animation is included with the online material.}
\end{figure*}

The high spatial resolution of the {\it IRIS} slit jaw imager allows us to examine the initiation of the eruption in unprecedented detail.  Figure \ref{triggering.fig} shows the initiation of the eruption in the {\it IRIS} 1330 \AA\ SJI as well as the AIA 304 \AA, 171 \AA, and 131 \AA\ channels.  The first indication that the prominence is destabilizing  in IRIS is the brightening of an elongated structure at 01:25:43 UT, indicated by a box in the top row of Figure \ref{triggering.fig}.  There is no indication of a brightening in the AIA 304 \AA\ or 171 \AA\ images at this time, and only a very weak brightening is seen in the center of the absorbing prominence material in the AIA 131\AA\ image.  A few minutes later, at 01:28:38 UT, the bright structure seen in IRIS futher stretches to the southwest.  At this time, there is  still no obvious counterpart to this structure in the AIA 304 \AA\ image, but both the AIA 171 \AA\ and 131 \AA\ images show a bright dot at the center of the absorptive material of the prominence, formed at the same time as the elongation of the bright feature in the {\it IRIS} SJI image.  Over the next minute, the elongated feature in the {\it IRIS} images splits in two, and the bright dot in the AIA images moves to the southwest in the images.  After this splitting, the prominence starts elongating in the same direction as the motion of the bright dot (see the online animation that accompanies Figure \ref{triggering.fig}).

The next phase of the eruption initiation involves the overlying material.  Tendrils of material from the overlying coronal rain structure can be seen interacting with the prominence in the {\it IRIS} SJI images as early as 01:30:54 UT.  In Figure \ref{triggering.fig}, we show images from 01:32:32 UT in {\it IRIS}, where the tendrils are very clear.  The tendrils connect to a brightening on the north side of the prominence that also appears in the AIA channels.  The tendrils are also visible in the AIA 304 \AA\ image, but because of the lower spatial resolution, the connections between the tendrils and the brightening are not as clear as they are in the {\it IRIS} images.  The tendrils are not visible at all in the AIA 171 \AA\ and 131 \AA\ images.  The bottom row of Figure \ref{triggering.fig} shows the early stages of the eruption at 01:36:25 UT.  The brightening seen previously at 01:32:32 UT has elongated, and at the end of the elongated bright structure there is a round ball of plasma that could be the erupting flux rope.  This structure is reminiscent of a flux rope and current sheet-like structure observed previously in the AIA 131 \AA\ channel during the well-studied eruption on 3 November 2010 \citep{Reeves2011, Foullon2011, Cheng2011, Bain2012, Savage2012b, Glesener2013, Hannah2013,Mulay2014}, although the plasma imaged here is much cooler, and the spatial extent of the eruption is smaller.

Because of the pre-existing structure that contains the coronal rain, it is tempting to look for signs of break-out reconnection, where the reconnection that triggers the eruption happens above the prominence \citep[i.e.][]{Antiochos1999, MacNeice2004,Lynch2008}, as was observed in a recent eruption studied by \citet{Sun2013}.  However, in this case, the observational features that can be linked to the eruption trigger seem not to occur above the prominence.  The separation of the elongated structure seen in the {\it IRIS} images between  01:28:38 UT and 01:29:37 UT could be due to reconnecting field lines separating the feature into two domains along the prominence.  Similarly, though the interaction with the tendrils of coronal rain does involve an overlying structure, the tendrils interact with the prominence itself low down, below the ball of material that perhaps constitutes an erupting flux rope.  This mechanism is consistent with some form of tether-cutting trigger \citep[i.e.][]{Moore2001,Sterling2005}.
 
Since many eruptions have a slow rise phase followed by a faster eruptive phase \citep[e.g.][]{Zhang2001,Sterling2005,Chifor2006}, we analyze the kinematics of the eruption in order to better understand the role of the brightenings shown in Figure \ref{triggering.fig}.  We trace the leading edge of the prominence to infer its bulk motion using a process similar to that described in \S\ref{aia_flows}. The data are first aligned using cross correlation to mitigate an orbital pointing wobble. A linear slice is then chosen to track the eruption using part of the prominence that is relatively free of flows that do not reflect  the bulk outward motion, namely those visible at the southern extent. This slice is shown in Panel a of Figure~\ref{fig:height_time} and is perpendicular to the limb with a width of two pixels.  Emission along the slice at a given time is stacked against subsequent observations to produce the height-time profile in Panel b, which is processed using the Canny edge detection algorithm to extract the leading edge. 

Pixels returned by the edge detection serve as our individual height measurements and are fit with the analytic approximation presented by \citet{Cheng2013} for their study of an active region flux rope eruption:

\begin{equation}  \label{eq-fit}
h(t) = c_0 e^{(t-t_0)/\tau} + c_1 (t - t_0) + c_2
\end{equation}

\noindent $h(t)$ is height, $t$ is time, and $\tau$, $t_0$, $c_0$, $c_1$, and $c_2$ are free parameters. This model combines a linear equation to treat the slow-rise phase and an exponential to treat the fast-rise. The onset of the fast-rise phase can be defined as the point at which the exponential component of the velocity equals the linear, which occurs at:

\begin{equation}  \label{eq-onset}
t_{\rm onset} = \tau \rm{ln}(c_1 \tau / c_0) + t_0
\end{equation}

Fitting is accomplished using MPFIT, a non-linear least squares curve fitting package for IDL \citep{Markwardt2009}.  Recently, we have used this method to examine the kinematics of 106 events observed by AIA \citep{McCauley2015}, and more detail on this method can be found in that work.

Using the above method, we find that the slow-rise phase lasts for $\sim$1.2 hours, beginning with an apparent velocity of  0.40 $\pm{}$ 0.05 km s$^{-1}$ before reaching twice that at $t_{\rm onset}$.  Because this velocity is so slow, we examine the effect of solar rotation, and find that up to half of the apparent velocity can be attributed to line of sight effects from the rotation of the prominence towards the limb, rather than an actual rise of the prominence above the solar surface.   After $t_{\rm onset}$, a rapid acceleration brings the final velocity to 250 $\pm{}$ 8 km s$^{-1}$ over $\sim$15 minutes.  We find the fast-rise onset time to be 01:25 UT $\pm{}$ 30s at a height of 7.3 $\pm{}$ 0.9 Mm above the limb. This time is very close to the time that the first brightenings are seen in the IRIS 1330 \AA\ SJI, as shown in Figure \ref{triggering.fig}.  Subsequent brightenings seen in IRIS and detailed in Figure \ref{triggering.fig} occur during the ramp up of the fast rise phase, and their times are marked with dotted lines in Figure \ref{fig:height_time}.

\section{Discussion and Conclusions \label{discussion.sec}}

  \begin{figure*}
 \includegraphics[scale=0.45]{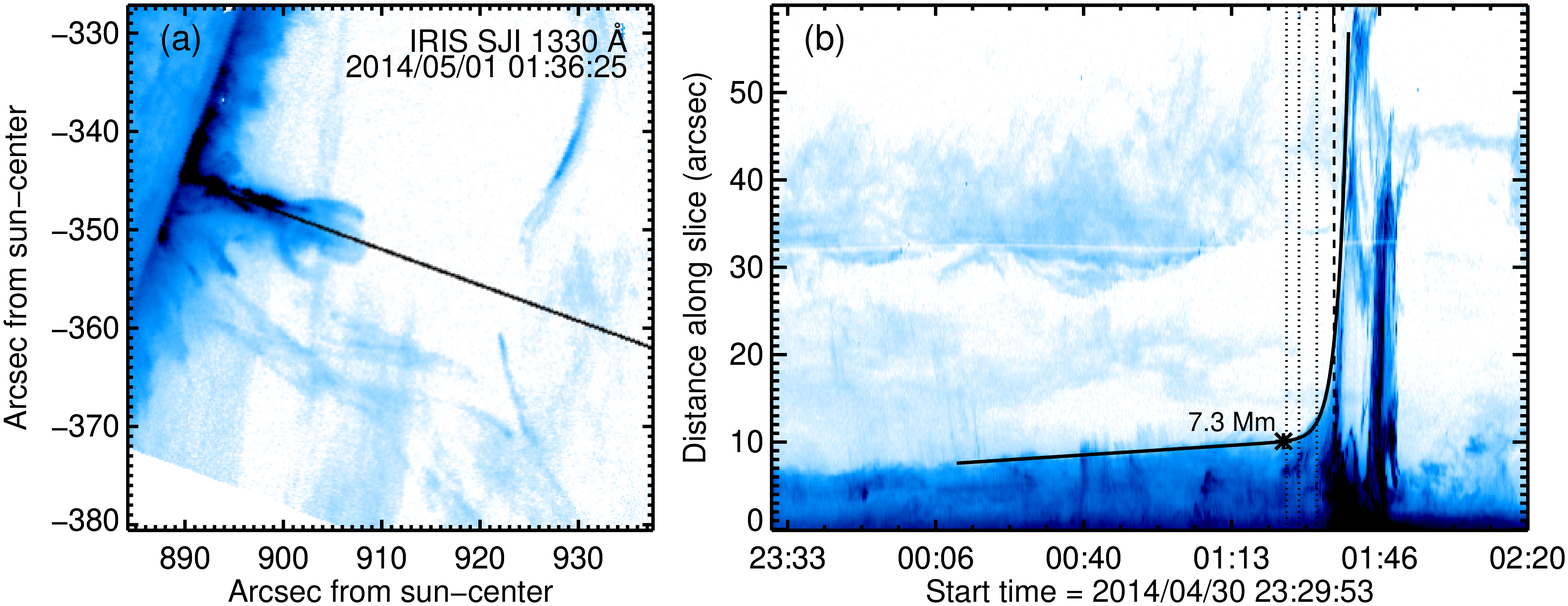}
  \caption{ Panel a) Slice used for tracking. Panel b) Height-time profile. The asterisk marks the fast-rise  onset, the dashed line indicates the time shown in Panel a, and the dotted lines indicate the timing of brightenings shown in Figure \ref{triggering.fig} at  01:25:43, 01:28:32, and 01:32:32 UT. See the online material for a corresponding movie.}
 \label{fig:height_time}
 \end{figure*}

\begin{figure}
\includegraphics[scale=0.4]{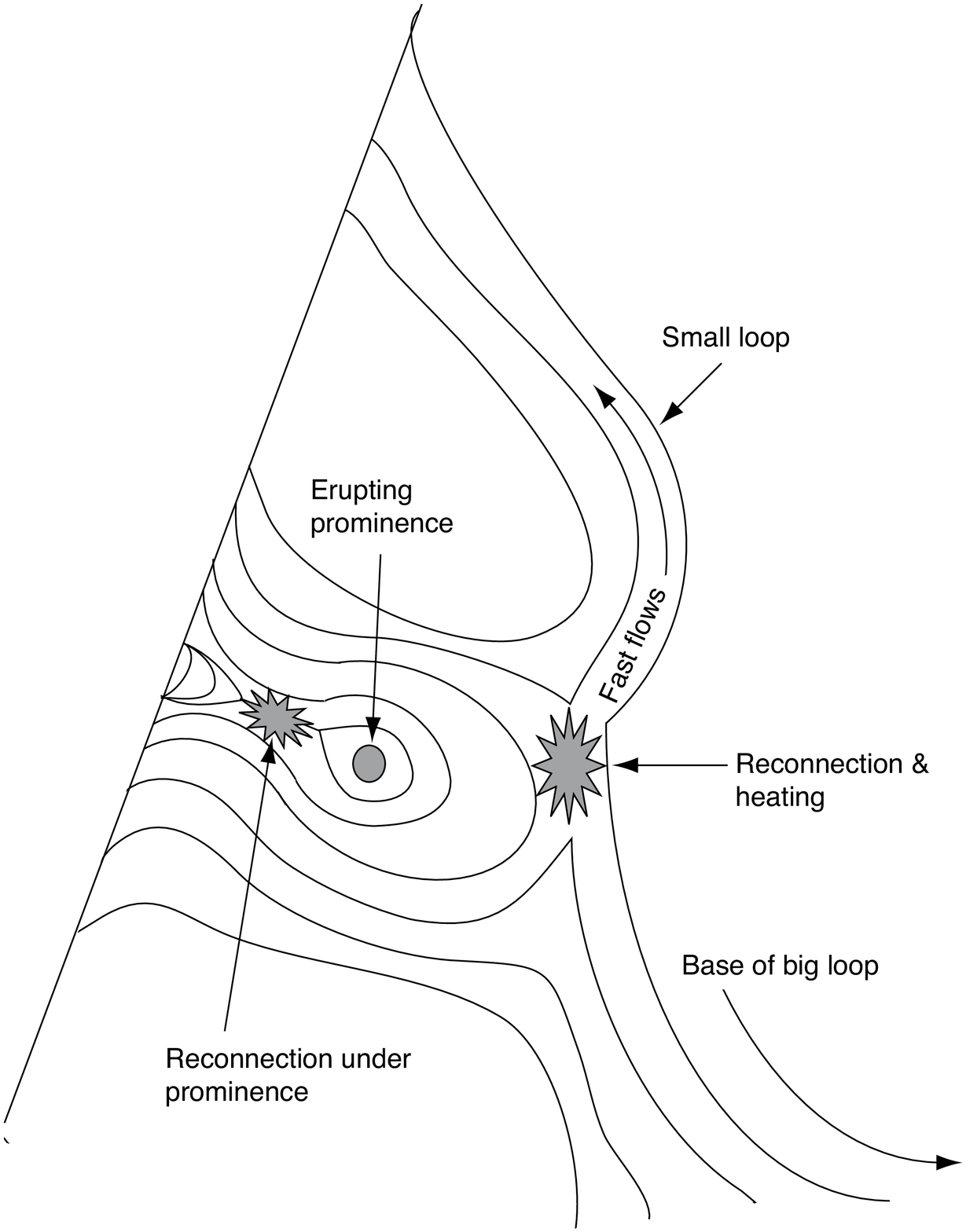}
\caption{\label{cartoon.fig} A cartoon showing the suggested magnetic configuration during the eruption.  Examination of the IRIS SJI images indicates that the reconnection below the prominence occurs before the reconnection above the prominence (see Figure \ref{triggering.fig}).}
\end{figure}

Taken as a whole, we interpret our observations to be evidence that reconnection is occurring between the prominence magnetic field and the the overlying magnetic field as the eruption takes place.  An illustration of the suggested magnetic field configuration is presented in the cartoon shown in Figure \ref{cartoon.fig}.  Additionally, we conclude that the reconnection is triggered not by break-out reconnection, but by reconnection occurring along and underneath the prominence.   The observational elements that lead us to these conclusions are as follows:
\begin{enumerate}
\item The kinked morphology of the overlying plasma as the eruption occurs, as seen in the {\it IRIS} SJI image in the second row of Figure \ref{summary.fig}.
\item The fast flows observed in both the {\it IRIS} Si IV line (Figure \ref{iris_redshift.fig}) and the AIA 171 \AA\ images (Figure \ref{AIA-velocity.fig}). 
\item Heating that occurs at the location of the kinked structure (which is itself the origin of the fast flows seen in AIA), as evidenced by hot plasma appearing in the AIA 131 \AA\ channel and the XRT Be-thin filter (Figures \ref{dem_small_loop.fig} and \ref{dem_big_loop.fig}). 
\item The eruption has a very slow initial rise phase ($\sim$0.4 km s$^{-1}$) that transitions into a very rapid acceleration and eruption with a final velocity of $\sim$250 km s$^{-1}$ (Figure \ref{fig:height_time}).
\item High spatial resolution {\it IRIS} SJI images   show brightenings at the time of the transition between the slow and fast rise phases, indicating possible locations for reconnections that trigger and drive the fast part of the eruption (Figure \ref{triggering.fig}).
\end{enumerate}

The line-of-sight fast flows we observe in {\it IRIS} using the Doppler shift of the {\it IRIS} Si IV 393.73 \AA\ line are intermittent, appearing and disappearing over the course of tens of seconds.  On the other hand, the flow seen in the AIA 171 \AA\ images is fairly steady.  We interpret the intermittency in the {\it IRIS} fast flows as being due to pre-existing cool material becoming entrained in the hotter fast reconnection flow seen in AIA.    The fast line-of-sight flows observed in {\it IRIS} spectra range from 112 -- 197 km s$^{-1}$, and the plane-of-sky flows observed in the AIA 171 \AA\ channel are about 300 km s$^{-1}$, giving a total velocity of 320 -- 360 km s$^{-1}$.  Other observations have found similar numbers for reconnection outflows.  \citet{Hara2011} found outflows of 200-400 km s$^{-1}$ in a small B9.5 flare observed with the Extreme ultraviolet Imaging Spectrometer (EIS) on {\it Hinode}.  Recently, \citet{Tian2014} found redshifts in the {\it IRIS} Fe XXI line indicating reconnection outflows with speeds up to $\sim$ 300 km s$^{-1}$ along the line of sight.  Supra-arcade downflows, thought to be related to reconnection outflows, typically have speeds on the order of several hundreds of km s$^{-1}$ \citep{Savage2011}.   

The pre-existing coronal rain in the area of the eruption has speeds of 65 km s$^{-1}$ and non-thermal widths of about 36 km s$^{-1}$.  Since the average non-thermal width of the Si IV line in the quiet Sun is about 20 km s$^{-1}$ \citep{TianScience2014}, these numbers indicate that the plasma in the overlying dome structure is dynamic and possibly turbulent even before the eruption.  During the eruption, the non-thermal widths get even higher, into the 50 km s$^{-1}$ range at Location 1, which is on the edge of the outflow that faces the overlying field.  Interestingly, the non-thermal width on the edge of the flow away from the overlying field (Location 2) has a very narrow non-thermal width of 23 km s$^{-1}$, on par with quiet sun values.  The difference in the non-thermal widths on different edges of the outflow could indicate small-scale plasmoid formation at the current sheet interface between the overlying field and the prominence field \citep[i.e.][]{Shen2011,Lynch2014}, or it could be indicitave of higher pressure underneath the dome structure, which would lead to compression that would impede the flow and narrow the line widths, as was seen by \citet{Harra2014}.

We find that at the locations of the fast flows observed by IRIS and AIA, there is a strong component of the DEMs at about 2.5 MK, about the same temperature as the background corona.  The DEMs indicate that there may be a small amount of hotter plasma as well, but the evidence is not definitive.  There is also a cool component to the fast flows, since they are visible intermittantly in the IRIS Si IV line.   This result seems surprising, since the reconnection process should heat the plasma as well as accelerating it.  However, if the magnetic field of the prominence is rotated with respect to the overlying field, a rotational discontinuity would form as the two fields reconnect.  The thermal properties remain constant across a rotational discontinuty, converting magnetic energy into purely kinetic energy \citep{PriestForbesBook}, which could explain the fast flows that do not appear to be heated early in the eruption.  Simulations of reconnection in skewed magnetic fields that include conduction show that the rotational discontinuity is followed by a slow shock that heats the plasma on timescales on the order of minutes \citep{Guidoni2010,LongcopeBradshaw2010}.  We do see clear evidence for heating at the reconnection site, given the emission that is seen in the AIA 131 \AA\ channel a few minutes after the flows are observed.  Unfortunately, the {\it IRIS} slit is not coincident with this location, so it is difficult to determine if this heating is due to shocks or not.  

The height-time plot in Figure \ref{fig:height_time} shows that the prominence undergoes a very slow rise phase with an apparent velocity of about 0.4 km s$^{-1}$, and the actual rise velocity during this phase could be somewhat lower due to line-of-sight effects.  This velocity is much slower than the typical slow rise phase velocity of an active region filament, which is usually on the order of 2-25 km s$^{-1}$ \citep{Sterling2005, Chifor2006, Cheng2013,McCauley2015}, although filaments in the quiet sun and polar crown filaments sometimes show slow rise velocities on the order of 1 km s$^{-1}$ or less \citep{Sterling2003,Sterling2004,Isobe2007,Regnier2011,McCauley2015}.  The only observation we are aware of with such a slow rise phase velocity in an active region filament is from work by \citet{LiuR2012}, who observe a three-stage filament eruption that starts with a very slow rise phase of 0.1 km s$^{-1}$, followed by a slow rise of 1 km s$^{-1}$, and then finally culminates in the rapid acceleration of the filament.  Active region eruptions often exhibit no slow-rise phase at all and may instead be fit with a single exponential \citep[e.g.][]{Williams2005}. These events may have slow-rise phases that are simply not perceptible by the instrument resolution or analysis technique. The fact that we can see such a small slow-rise velocity in such a small structure may be attributable to the fine spatial resolution of IRIS.  Some possible mechanisms for the slow rise phase are flux cancellation that  forms the filament and increases its twist \citep[e.g.][]{Amari2010,Aulanier2010}, or emergence of the flux rope from beneath the photosphere \citep{Fan2007}, though because this filament is on the limb, it is difficult to say which of these mechanisms might be occuring.
  
The high spatial resolution images from the {\it IRIS} slit jaw imager allow us to examine the triggering of the fast rise in great detail. We find that there is a brightening in IRIS at about the same time as our calculated switch over time from the slow rise phase to the fast rise phase.  A very faint indication of this feature is seen in the AIA 131 \AA\ image, but there is no sign of it in the 171 \AA\ image until a few minutes later, when a bright dot appears in the center of the prominence. Tendrils of material associated with the overlying field are clearly seen in the IRIS images as the fast rise phase begins, and they seem to interact with the prominence,  possibly contributing further to destabilizing it.  Similar features can be seen in the AIA 304 \AA\ channel, but it is difficult to follow their motion due to the lower spatial resolution.  The coronal wavelengths (i.e. 171 and 131 \AA) show completely different features during the eruption triggering that again are not entirely clear due to the spatial resolution.  A high resolution EUV telescope, similar to the one flown on the Hi-C rocket \citep{Kobayashi2014}, would be extremely beneficial for future studies of the triggering mechanisms in coronal mass ejections.

\section*{Acknowledgements} 
The authors would like to thank the anonymous referee for thoughtful comments that improved this paper, and Patrick Antolin, Ed DeLuca, Wei Liu, Nick Murphy, and Xudong Sun for helpful discussions.  This work is partially supported by contracts 8100002705 and SP02H1701R from Lockheed-Martin to SAO and contract NNM07AB07C from NASA to SAO.   K. Reeves and P. McCauley also acknowledge support from NASA grant NNX12AI30G to SAO.  {\it IRIS} is a NASA small explorer mission developed and operated by LMSAL with mission operations executed at NASA Ames Research center and major contributions to downlink communications funded by the Norwegian Space Center (NSC, Norway) through an ESA PRODEX contract.  {\it Hinode} is a Japanese mission developed and launched by ISAS/JAXA, with NAOJ as domestic partner and NASA and STFC (UK) as international partners. It is operated by these agencies in co-operation with ESA and NSC (Norway).  This work has benefited from the use of NASA's Astrophysics Data System.

\noindent{\it Facilities}: {\it Hinode} (XRT), {\it SDO} (AIA), {\it IRIS}

\bibliographystyle{apj}

\end{document}